# Exploring the Structure and Chemistry of 1D and 2D Lepidocrocite TiO$_2$ at Atomic Resolution

*Eric Nestor Tseng,[a] Jonas Björk,[b,c] Risha Achaiah Iythichanda,[a,c] Wei Zheng,[b] Jie Zhou,[b] Hatim Alnoor,[a] Wei Hsiang Huang,[d,e] Ming-Hsien Lin,[f] Johanna Rosen,[b,c] Per O.Å. Persson.[a,c]\**

[a] Thin Film Physics Division, Department of Physics, Chemistry and Biology (IFM), Linköping University, 58183 Linköping, Sweden

[b] Materials Design Division, Department of Physics, Chemistry and Biology (IFM), Linköping University, 58183 Linköping, Sweden

[c] Wallenberg Initiative Materials Science for Sustainability (WISE), Linköping University, Department of Physics, Chemistry and Biology (IFM), 58183 Linköping, Sweden

[d] National Synchrotron Radiation Research Center (NSRRC), Hsinchu City, 30076, Taiwan

[e] Sustainable Electrochemical Energy Development (SEED) Center, National Taiwan University of Science and Technology, Taipei 106, Taiwan

[f] Department of Chemical and Materials Engineering, Chung Cheng Institute of Technology, National Defense University, Taoyuan, Taiwan

*Corresponding author:
per.persson@liu.se

**ABSTRACT**

Low-dimensional materials are critical for enabling next-generation applications that are central to addressing critical global challenges. Titanium dioxide (TiO$_2$) nanostructures stand out due to their structural versatility and relevance to catalysis, energy conversion, and environmental remediation. Here, we employ a combination of advanced electron microscopy, spectroscopy, and first-principles theoretical calculations to investigate the structural and chemical properties of one- and two-dimensional lepidocrocite-type TiO$_2$. Special emphasis is placed on the one-dimensional material, which exhibits anisotropic growth, extending exclusively along a single crystallographic direction. Our analysis suggests that this unusual growth behavior can be attributed to light-element impurities, such as carbon, that are incorporated during the bottom-up synthesis. The results extend the understanding for these unexplored low-dimensional TiO$_2$ materials and offer fundamental insights into their structure and chemistry.

**INTRODUCTION**

Since the discovery of graphene,[1] research into other two-dimensional (2D) materials such as boron nitride,[2] transition metal dichalcogenides [3] and MXenes[4] has increased significantly. Interest in reducing the dimensionality of materials is manifold. First, when materials are confined by reduced dimensionality, quantum confinement leads to physical properties that are absent in bulk. Beyond these changes, the surface area normalized by weight or volume increases dramatically with decreased dimensionality. This is essential for applications in which interactions between surface and the environment is critical, such as energy storage, catalysis, filtering, and capture.

Titanium dioxide ($TiO_2$) nanostructures have attracted significant attention due to unique properties, that makes them appealing for a wide range of applications, ranging from paint pigment, (photo-)catalysis, sensors, solar and fuel cells.[5-15] While most $TiO_2$ nanostructures are 3D particles of nanoscale size, atomically thin structures have also been reported. Recently, two $TiO_2$ based low-dimensional materials were independently demonstrated, exhibiting 2D[16] and one-dimensional (1D)[17] morphologies, respectively. The 2D structure was synthesized through a top-down prosses, using molten salt etching of a layered Ti-based boride ($Ti_4MoSiB_2$), while in contrast the 1D structure was obtained through a bottom-up process using TMAOH etching of TiC. The structures are morphologically distinct: the 2D material, like other 2D structures, form sheets one unit cell thick, whereas the 1D material exhibits a cotton-like morphology associated with high permeability.

In the present work, we show that both methods result in the same atomically thin lepidocrocite structure. In the 2D case the material forms sheets with lateral dimensions in the micrometer range, whereas the 1D form consists of filaments with widths of only a few nanometers and lengths exceeding hundreds of nanometers, beyond the experimentally accessible range. Although lepidocrocite $TiO_2$ is a known layered structure, it has not previously been synthesized in a strict 2D form. Herein, both low-dimensional forms are investigated using atomically resolved scanning transmission electron microscopy (STEM), electron energy loss spectroscopy, X-ray absorption spectroscopy, and first-principles calculations. Together, these approaches provide a comprehensive understanding for the structure and chemistry of these low-dimensional materials, including defects and stoichiometry.

**RESULT AND DISCUSSION**

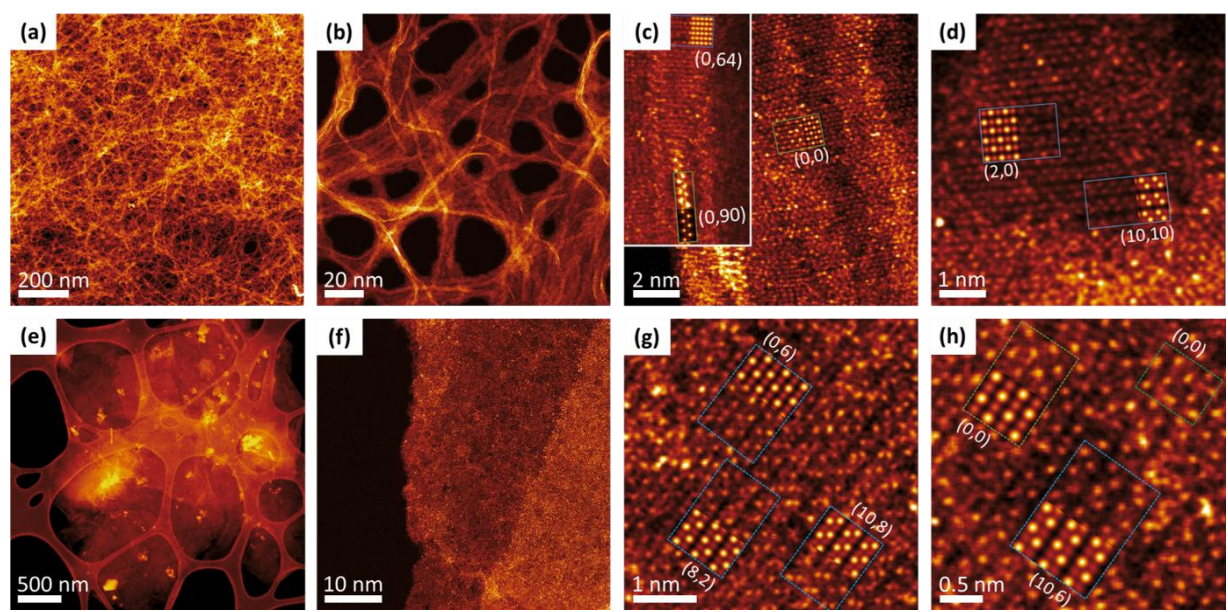

**Figure 1.** *Low magnification (a, b, e, and f) and atomic resolution (c, d, g and h) STEM images of 1D (top row) and 2D (bottom row) lepidocrocite titania, respectively. The inset in c) verifies the unit cell thin structure in cross-section. The overlays in the atomic resolution images are simulated images of the Lepidocrocite structure, while the associated numbers indicate applied tilt (°) in x and y.*

Sample solutions containing the 1D and 2D material were drop casted on lacey carbon film in gold grids. They were subsequently investigated by HAADF-STEM and the results are shown in Figure 1. The one-dimensional material results are shown in Figure 1 (a-d), where Figure 1 (a) shows an overview image of the sample, presenting a cotton-like microstructure that appears to exhibit both exceptional surface area and permeability. At higher magnification, Figure 1(b), the material is observed to be composed from nanoscale filaments of undetermined length, that are exhibiting random orientation and pronounced curvature. The bright lines in Figure 1 (b) appear when filaments are oriented with the edge towards the electron beam, thereby increasing the projected thickness, while otherwise oriented they appear dull in comparison. This infers that the filaments are exceptionally thin. Figures 1 (c-d) show the filaments at atomic resolution and verify the atomically thin structure (see inset in Figure 1c). When oriented in plan-view it is clear that the filaments exhibit a varying width of 3-6 nm, where the edges of the otherwise structurally ordered filament are disordered. Moreover, Figures 1 (c-d) demonstrate the atomic arrangement of the filament, enabling structural characterization.

The 2D sample was similarly prepared and investigated by HAADF-STEM, with the corresponding results shown in Figures 1 (e-h). Figure 1 (e) displays agglomeration of 2D flakes, where the 2D nature is clearly identified by the integer variation of image intensity associated with increasing number of overlapping flakes, at the edge of the agglomerate, see Figure 1 (f). Figures 1 (g-f) reveal the atomically resolved structure of the 2D sheets, that display a structural organization that is identical to the 1D filaments.

To further explore the structure and the composition, both materials were examined using EELS. Core-loss spectra and corresponding quantification results are shown in Figure S1. From the results, the materials both consist of titanium and oxygen in a relative composition of 1:2, suggesting titania. Possible titania structures were therefore considered, where the most common (e.g. rutile and anatase) can be ruled out based on lattice spacings and symmetry. Instead, Lepidocrocite titania can be matched to the structures observed by STEM, wherein $TiO_6$ octahedra are arranged in two dimensions.

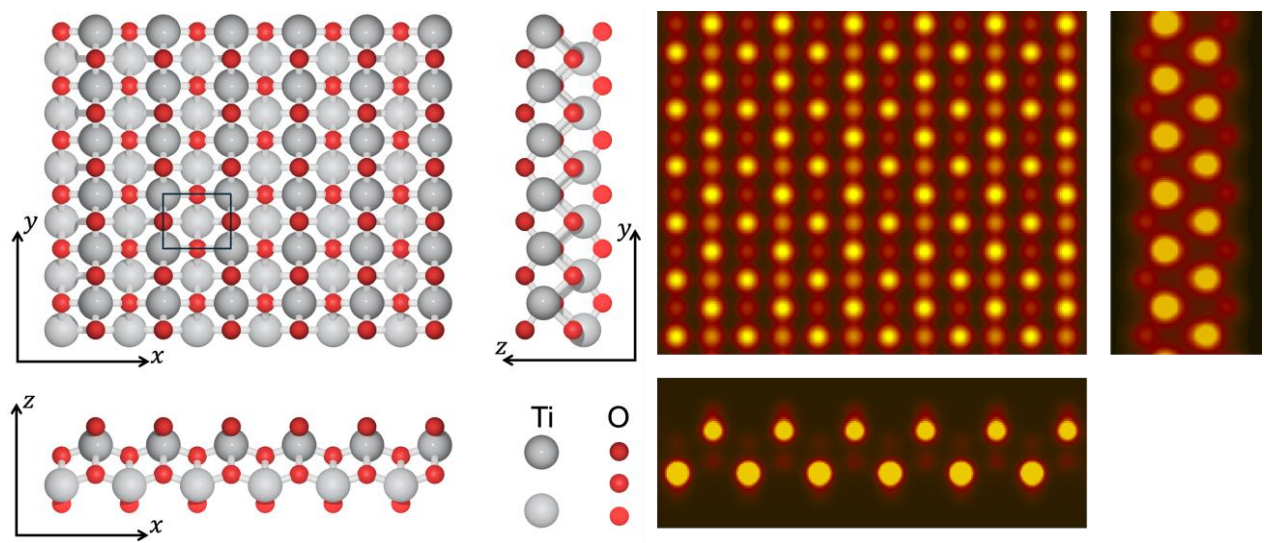

**Figure 2.** *Ball and stick models of a single sheet of lepidocrocite titania (left) in to-view (x-y) as well its cross section (x-z, y-z). The corresponding simulated STEM images of the single lepidocrocite sheet are shown (right).*

Figure 2a shows the structure of a single $TiO_2$ lepidocrocite sheet in top view (x-y), and in two inequivalent cross sections (x-z, y-z). Note that the unit cell (indicated in the top view structure) is rectangular, and therefore the appearance of the top view structure also becomes rectangular, although there are as many atoms in the x-direction as in the y-direction. Both cross-sections reveal an undulating appearance. STEM images of these structures were simulated as shown in Figure 2b. The Ti atoms stand out due to their heavier mass, but the for the top view image the superimposed contrast from two O atoms is also clearly visible. In cross section the contrast from Ti is substantially stronger than from O. However, comparing with the acquired high-magnification images in Figure 1, it is apparent that the simulated images does not perfectly match, as apparent "gaps" are visible in the acquired images.

Figure S2 display simulated STEM images of 2D Lepidocrocite titania for different axis projections and tilt in *x* and *y*, respectively. Interestingly, the tilt series reveal how an apparent gap appears in the image between the atoms, when the structure is rotated in the x-direction, though notably not in the y-direction. This suggests that the discrepancy between the simulated image ion Figure 2 and the experimentally acquired images in Figure 1, owes to sample tilt. However, results reveal that this gap is always aligned perpendicular to the filament growth direction, which can be used to quickly identify that the filaments are exclusively extended in the x-direction.

Overlays of the simulated structure at different crystal rotations (indicated at the bottom of the overlay) are inserted in Figures 1 c-d, e-f to match the underlying structure. Remarkably, the structure of both materials can be bent by approximately 10°, over a distance of <5 nm (see e.g., Figure 1d), suggesting that the structures are subject to substantial stress or are structurally modified through point defects. To support this, vacancies can be observed both on the Ti and on the O sites in the atomically resolved images. Single Ti atoms, presumably residual atoms originating from synthesis, can be observed to decorate the surfaces together with lighter elements that can be O atoms or contamination (such as C).

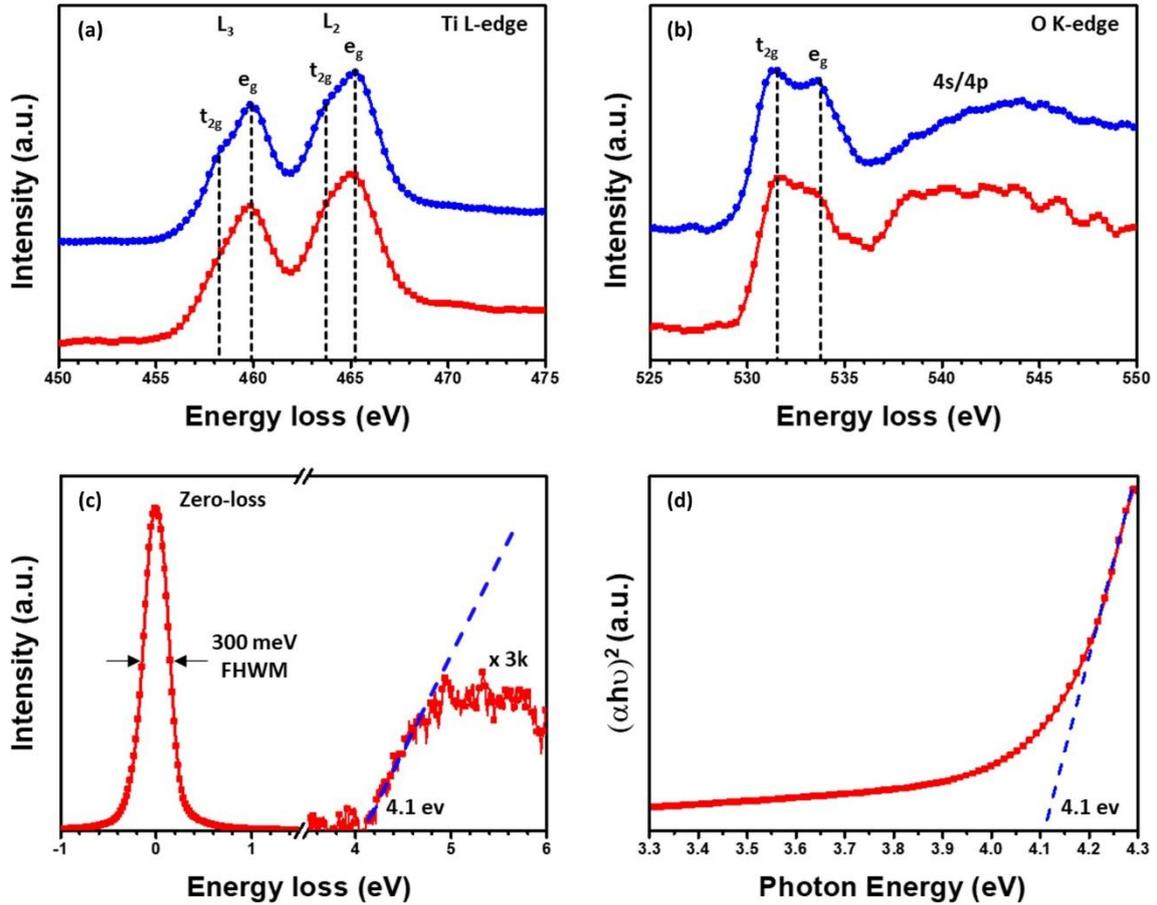

**Figure 3.** *Electron energy loss spectra in for the Ti-$L_{3,2}$ (a) and O-K (b) -edges from the one- and two-dimensional Lepidocrocite in red and blue, respectively. A monochromated valence electron energy loss spectrum from the one-dimensional Lepidocrocite is shown in (c) and a Tauc plot of UV-Vis spectra in (d).*

To further explore the chemistry of these low-dimensional lepidocrocite materials, fine structure EELS analysis was performed of both materials. In Figure 3a, the core-loss EELS spectra of the Ti-$L_{3,2}$ edges for the 1D and 2D materials are shown in red and blue, respectively. The spectra reveal two main peaks for both materials, where a distinct field splitting is observed in the spectrum acquired from the 2D material, while it is visible but comparably less pronounced in the spectrum acquired from the 1D material.

The Ti-L edge arises from Ti 2p core electron excitation into Ti 3d and Ti 4s unoccupied states. Due to spin-orbit splitting, the separating energy between 2p1/2 and 2p3/2 core holes give rise to the $L_2$ and $L_3$ edges, respectively.[21] As observed, apart from the spin-orbit splitting of 2p states, the 2D material is additionally split into $t_{2g}$ and $e_g$ states. The observed split into additional peaks may depend on various items, including oxidation state, as well as the coordination and site symmetry.[22] These four pronounced features are commonly observed among $Ti^{4+}$ compounds with $TiO_6$ coordination. The 1D Lepidocrocite is observed to exhibit comparably weak $t_{2g}$ intensity, which may be caused by a high amount of d electrons occupying the outer Ti orbital. In extension, this suggests O vacancies and/ or a distorted O sublattice. This agrees with the STEM observations made in Figure 1, which reveal both point defects and a strained structure. It is also observed that between the two materials, there is no chemical shift between the edges, which suggest the same or similar bonding state. However, for both the 1D and 2D Lepidocrocite, the oxidation state for Ti is judged as tetravalent.

The O-K edge, as shown in Figure 3b reflects the core electron transitions from O 1s to O 2p unoccupied states, and exhibits two pronounced peaks. These two initial peaks reflect the hybridization between the O 2p and the Ti 3d orbitals and are accordingly split into $t_{2g}$ and $e_g$, respectively. At higher energies, a broad convolution of peaks is formed through O 2p hybridized with Ti 4s and Ti 4p states.[23] It may be observed that the 1D material again reveals a less pronounced splitting between $t_{2g}$ and $e_g$, which again may be inferred to occur because the 1D material experiences a more defective and distorted lattice compared to the two-dimensional material. Similar to the Ti $L_{3,2}$ edge, no chemical shift can be observed, which emphasizes the identical bonding of the two materials.

A monochromated electron beam was utilized to probe the VEELS properties of the 1D material, and the results are shown in Figure 3c. It is apparent that the material exhibits a band gap, $E_g$, reflected by the broad intensity increase which is generated by electron transitions from the valence band to the conduction band.[24] The bandgap is estimated to 4.1 eV, on par with a previous measurement of the 2D material.[16] While the VEELS measurement probes a very small volume, a corresponding UV-vis spectroscopy measurement was performed on the 1D material and the results are shown in Figure 3d, where the Tauc plot corroborates the 4.1 eV bandgap obtained by VEELS. The bandgaps for both the 1D and 2D lepidocrocite materials exhibit a significantly larger value compared with other $TiO_2$ based materials and likely relates to the low dimensionality of the structure.[25]

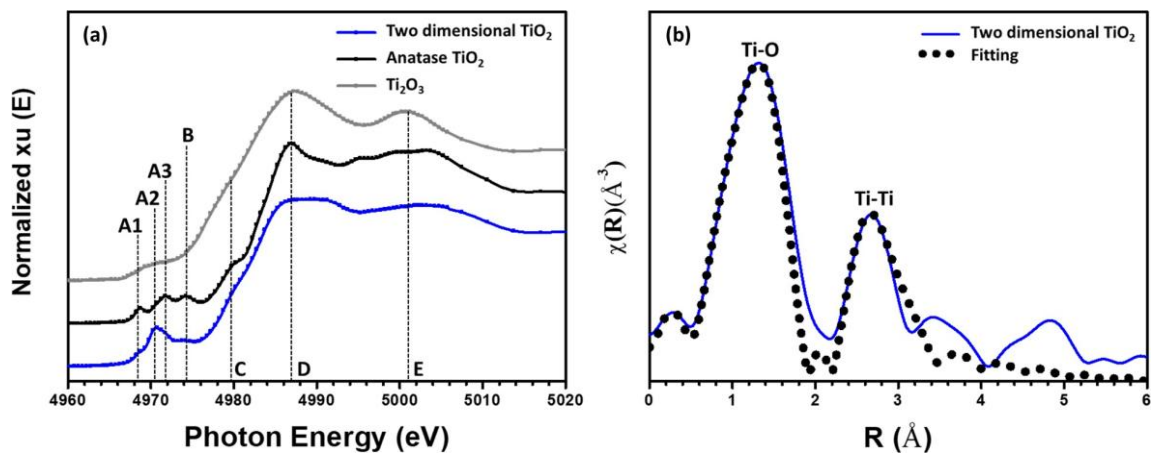

**Figure 4.** *a) X-ray absorption near edge spectra from the Ti K-edge of the 2D Lepidocrocite compared with Anatase titania and $Ti_2O_3$. Titania pre-peaks (A1, A2, A3 and B) and main peaks region (C, D and E) are indicated for reference. The absolute Fourier transforms of the extended fine structure region for two-dimensional Lepidocrocite, together with a fitting curve are shown in b).*

The 2D material was further explored by X-ray absorption spectroscopy with respect to local atomic structure and oxidation state using XANES and EXAFS of the Ti K-edge. As shown in Figure 4. The near edge structures of the 2D material are compared with Anatase titania and $Ti_2O_3$ in Figure 4a. For reference, Anatase titania shows three peaks in the pre-edge region, denoted A1, A2 and A3, respectively, which are assigned to dipole-forbidden 1s to 3d transitions.[26] These are highly sensitive to the coordination of the Ti site, indicating four-, five-, and six-fold coordination of the oxygen atoms, respectively. The 2D material spectrum demonstrates a significant A2 peak, which emphasizes a five-fold coordination for the Ti atoms. Such spectra are known to appear for bulk materials that are rich in oxygen vacancies.[21] The three main-edge peaks denoted C, D and E correspond to the 1s to 4p transitions make up the main shape of the edge and strongly correspond to previous reports on pristine layered titanate or lepidocrocite-type structure[27] and corroborate the atomically resolved STEM results. $TiO_2$ materials that are rich in O vacancies could potentially change the Ti valence from $Ti^{4+}$ to $Ti^{3+}$, however, when the 2D material spectrum is compared with the $Ti_2O_3$ reference structure, the resemblance is negligible.

We additionally performed EXAFS analysis, where the absolute Fourier transform of the 2D material spectrum and corresponding fit is shown in Figure 4b with the fitting parameters listed in Table S1. The analysis shows that the main shell of Ti-O and Ti-Ti atomic radius distance are 1.83 Å and 3.23 Å respectively. Consistent with the XANES region, the Ti-O bonding in the first shell with average coordination numbers is found to be 4.25, which is slightly less than perfect $TiO_2$ and presumably owing to the defective nature of the material, exhibiting vacancies.

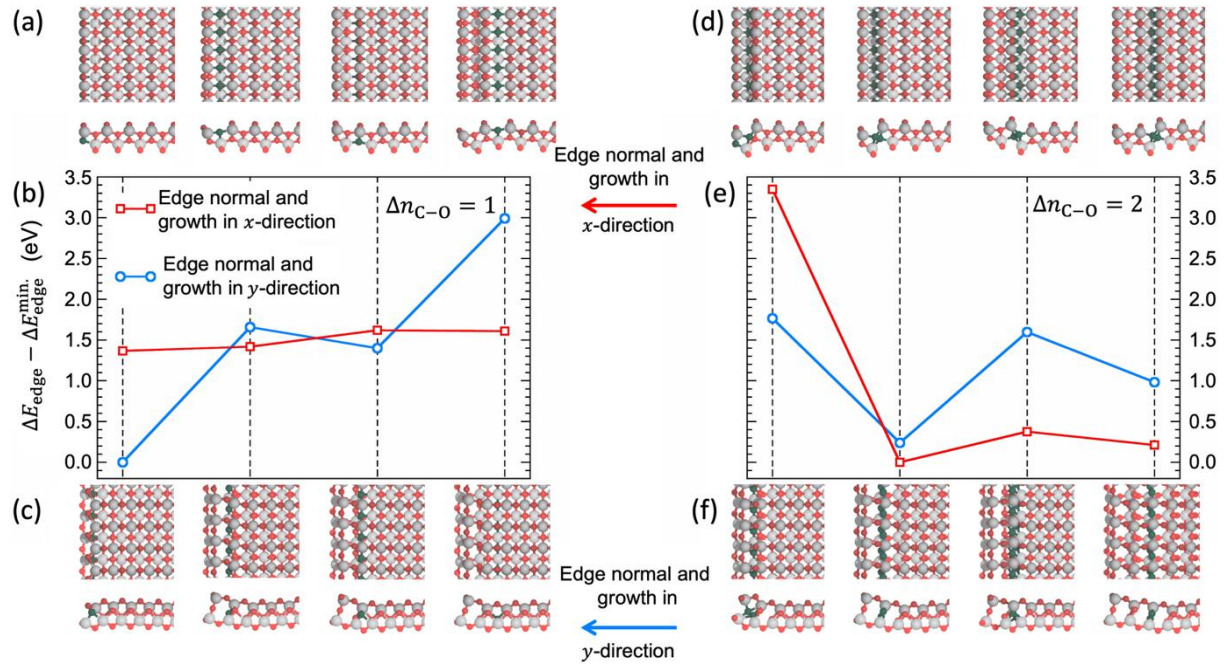

**Figure 5.** *Comparison of $TiO_2$ thin foils for which growth proceeds along the x-, and y-directions, where the edge normal is equivalent to the growth direction, containing oxygen-to-carbon substitutions at different distances from the edge. (a–c) Structures with one O atom substituted by one C atom per Ti edge atom: foils with growth progressing in the x-direction (a) and the y-direction (c), and the corresponding relative edge energies (b). (d–f) Structures with two O atoms substituted by two C atoms per Ti edge atom: foils with growth progressing in the x-direction (d) and the y-direction (f), and the corresponding relative edge energies (e). Relative energies in (b) and (e) are referenced to the most stable configuration for each composition and are therefore not directly comparable. Panels (a), (c), (d), and (f) show both top and cross-sectional views, focusing on one of the two equivalent edges of each filament. Ti, O, and C atoms are shown in gray, red, and dark green, respectively.*

Figure 1 shows that the one-dimensional material extends only in the $x$-direction, whereas the 2D-material extends in both directions. To get insight into the origin of this anisotropic growth behavior, we performed density functional theory (DFT) calculations on lepidocrocite foils with different edge structures and orientations. The stability of the different edge structures is assessed by their edge energies, defined as:

$$\Delta E_{\text{edge}} = \frac{1}{2n_{Ti}^{\parallel}} \left[ E_{\text{1D-Lepi}} - n_{Ti}^{\parallel} n_{Ti}^{\perp} E_{\text{2D-Lepi}} - 2\Delta n_{\text{C-O}} \Delta E(\text{C} - \text{O}) \right],$$

Here, $E_{\text{1D-Lepi}}$ is the total energy of the foil under consideration, and $E_{\text{2D-Lepi}}$ is the total energy per TiO$_2$ unit of the corresponding extended 2D lepidocrocite sheet. The integer $n_{Ti}^{\perp}$ specifies the foil width (number of Ti atoms across the filament), while $n_{Ti}^{\parallel}$ gives the supercell size along the periodic direction (perpendicular to the growth direction). In all calculations presented here, $n_{Ti}^{\parallel} = 1$. The normalization factor $2n_{Ti}^{\parallel}$ implies that the edge energy is given with respect to the number of Ti-atoms at each edge. Because the material is formed under carbon-rich conditions, oxygen-to-carbon substitutions are included explicitly. Here, $\Delta n_{\text{C-O}}$ is the number of O atoms replaced by C per Ti edge atom and $\Delta E(\text{C} - \text{O})$ is the reference energy for such a substitution. Importantly, when comparing edge energies for the same composition, this reference energy is cancelled.

The edge energy can be interpreted as the energy cost of creating an edge relative to the corresponding 2D sheet. However, for bottom-up growth processes, its physical meaning depends on the growth mechanism. Under thermodynamic control – i.e., when growth proceeds through reversible steps – the system evolves toward a shape that minimizes the total free energy. Growth therefore occurs preferentially normal to edges with higher edge energy, reducing their extent and leaving predominantly low-energy edges exposed, consistent with the principles underlying the Wulff construction.

In contrast, under kinetic control, where key reaction steps are irreversible, growth rates are controlled by activation energies rather than by thermodynamic stability. According to the Brønsted−Evans−Polanyi relation, activation energies scale approximately linearly with reaction energies for similar reactions. The reaction energy associated with widening a filament by one structural unit is $E_{n+1} - E_n - \mu$, where $\mu$ is the chemical potential of the growth species. This quantity is governed by the change in edge energy with increasing width, meaning that the edge energy serves as a proxy for both the reaction energy and the corresponding activation energy. An increasing edge energy with width implies progressively unfavorable widening, whereas a decreasing edge energy indicates energetically favorable growth. When the edge energies become width-independent – i.e., once the filament interior reaches the bulk-like limit

– the reaction energies, and thus the growth rates, depends only on the chemical potential of growth species in those directions, irrespective of which direction has the lower absolute edge energy.

With these arguments in mind, we first consider impurity-free foils of various widths extended along the two main crystallographic directions ($x$ and $y$), see Figure S3. The energy of an edge with its normal to the $y$-direction is generally higher than that for an edge with its normal along $x$. According to thermodynamic arguments based on the Wulff construction, growth would proceed preferentially along the $y$-direction. In this way, the relative length of high-energy edges is reduced, while low-energy edges are dominating, minimizing the total free energy. From a kinetic perspective, growth is initially faster along the $x$-direction, where the edge energy is roughly constant with width, making row addition equally favorable for narrow and wide foils. In contrast, edges that grow along $y$ exhibit an increasing edge energy for narrow widths, making early widening less favorable. As the foils broaden, edge energies in both directions become width-independent, and reaction energies – and thus growth rates – converge, leading to isotropic growth and formation of the equiaxed 2D sheets observed above in the absence of carbon.

Exclusive growth along the x-direction is observed experimentally for filaments that are formed when carbon is present during synthesis (vide supra), motivating an investigation on the effect of C incorporation on the edge energies. Figure 5 compares foils of width $n_{Ti}^{\perp} = 20$ along the x- and y-directions with different carbon content: either replacing one O atom with one C atom per Ti edge (Figure 5a-c), or two O atoms with two C atoms per Ti edge (Figure 5d-f). For each composition, the edge energies are given relative the corresponding most stable structure, ensuring that the reference energy associated with the O-to-C substitution is eliminated. Note that to compare the energies for different carbon content, this reference energy is still required.

For the low carbon content (Figure 5b), with the C-atom positioned at the edge, the foil with edges growing along $y$ exhibit a lower edge energy than that with edges growing along $x$. However, moving the C impurity toward the center of the foil significantly increases the edge energy for edges growing along $y$, while for edges growing along $x$ there is virtually no energy penalty for moving the C impurity toward the interior. This indicates that foils growing along the $x$-direction can extend without notably increasing the edge energy, even when C is present at the edge, making such growth kinetically accessible. In contrast, for growth along the $y$-direction, C atoms at the edge can effectively hinder further growth, as incorporating C into the interior of the TiO$_2$ filament becomes energetically unfavorable.

For the high carbon content (Figure 5e), incorporation of C atoms at the outermost edge is energetically unfavorable for both directions, although it is slightly more stable for edges growing along y. However, moving the C impurities toward the interior, is significantly easier for the edges growing along $x$, facilitating growth in this direction. This behavior likely originates from substantial edge reconstruction for edges growing along $y$, which alters the edge topology and makes the energy more dependent on the specific positions of the C atoms. Importantly, for both carbon cases, the edge energy is less sensitive to the position of the carbon impurities for edges growing along $x$, favoring growth and filament formation along this direction.

The microscopy results indicate that the filaments bend considerably. This can partly be explained by the microstructure which pins the filaments in continuously varying orientations, but bending also occurs in the vicinity of a substitutional impurity, as seen from the structures presented in Figure 5. The calculations may also serve to explain both the varying width of the filaments as impurities are incorporated at random during growth, and the disordered edges of the filaments, which compare well with the significant edge reconstruction that occurs for the high concentration impurity containing filament for growth along y. Despite challenges associated with directly resolving carbon impurities at the atomic scale, the combined experimental observations and theoretical results consistently support their critical role in directing anisotropic growth.

More broadly, the results establish a clear relationship between impurity incorporation, edge energetics, and growth dimensionality in lepidocrocite TiO$_2$. This insight not only explains the formation of the one-dimensional filaments but also suggests that dimensionality in similar systems may be tunable through deliberate chemical control during synthesis.

Put together, the observed structural flexibility, defect-rich nature, and tunable growth behavior underscore the potential of these materials as a platform for further tailoring of structure–property relationships in low-dimensional materials. The findings therefore open pathways for TiO₂-based nanostructures with tailored properties for applications where high surface area, permeability, and controlled electronic structure are essential.

**CONCLUSION**

1D filaments and 2D sheets of lepidocrocite structured TiO₂, constitute a new and exciting addition to the family of low-dimensional materials. These materials and in particular for the 1D filaments the microstructure offer a material with unprecedented surface area together with exceptional permeability. The present investigation reveals insights into the structure, chemistry, composition, defects and impurities of these materials. For the first time, a single lepidocrocite TiO₂ filament has been resolved in plan-view, revealing intrinsic defects and atomic-scale features. These observations infer carbon substitution for oxygen as a plausible mechanism underlying the strictly one-dimensional growth of the filaments.

**EXPERIMENTAL DETAILS**

Materials synthesis:

The 1D material was prepared in a similar fashion as described ealrlier[17], with a slight modification. 1 g of titanium carbide (TiC, 2 μm, Alfa Aesar) was immersed into 20 ml of tetramethylammonium hydroxide (TMAOH) in a polyethylene vial and heated to 80 °C in an oil bath while subjected to continuous magnetic stirring, at 500 rpm, for 2 days. At this point, a dark grey sediment was obtained and was washed several times in ultra-pure water using a centrifugation process, 5500 rpm for 5 minutes each time, until the solution reached a pH of 6.5–7. Subsequently, the collected sediment was re-dispersed in 45 ml of ultrapure water and sonicated in a water bath for 30 minutes. Finally, the dispersion was centrifuged at 3500 rpm for 15 minutes from which a stable colloidal suspension was obtained for further characterization.

The 2D material was obtained as described previously, from $ZnCl_2$ molten salt etching of the precursor $Ti_4MoSiB_2$.[16]

Materials Characterization:

Samples for (S)TEM were prepared by drop-coasting a 1 μl droplet of sample solution on a lacey carbon gold grid. High angle annular dark field STEM (HAADF-STEM) and valence electron energy-loss spectroscopy (VEELS) were obtained using the monochromated and double-corrected Linköping FEI Titan³ 60–300 kV and the embedded Gatan GIF Quantum ERS spectrometer operated at 300 kV. VEELS was recorded with an energy resolution of 300 meV at an accelerating voltage of 300 kV. The valence spectrum was recorded by rastering the defocused probe across a ~1 μm² area containing multiple filaments. The valence region was background subtracted using a standard power-law model embedded in Gatan Digital Micrograph.

X-ray absorption near edge structure (XANES) and extended fine structure (EXAFS) spectroscopy measurements were performed at National Synchrotron Radiation Research Center (NSRRC) in Taiwan, hard-XAS of the Ti K-edge was measured at the BL17C1 beamline. The experiments at the BL17C beamline were performed with a graphite filter, first slits, vertical collimating mirror (VCM), second slits, double crystal monochromator (DCM), toroidal focusing mirror and ionization chambers. The VCM mirror is made from Rh metal coated on the surface of a Si substrate. Two parallel Si (111) crystals of DCM were used for energy selection. All of the computer programs were implemented in the Athena package with the backscattering amplitude and the phase shift for the specific atom pairs being theoretically calculated by using Artemis code along with self-created model.

UV–vis spectra were obtained using a spectrophotometer (Lambda 900, Perkin Elmer Instruments). Measurements were performed in absorption mode, in the range of 200 to 800 nm, using a dilute solution of the sample into a quartz cuvette.

Computational Details:
Periodic density functional theory (DFT) calculations were performed using the Vienna *Ab initio* Simulation Package (VASP),[18] employing the projector-augmented wave (PAW) method[19] and a plane-wave basis set with a kinetic energy cutoff of 520 eV. Exchange–correlation effects were treated within the Perdew–Burke–Ernzerhof (PBE) functional.[20] The Brillouin zone was sampled using 11 *k*-points along the periodic direction of the one-dimensional structures and a single *k*-point in the non-periodic directions. A vacuum spacing of 14 Å was applied in the non-periodic directions to prevent interactions between periodic images. Structural relaxations were carried out until the residual forces on all atoms were below 0.02 eV Å⁻¹, while the lattice constant along the periodic direction was fixed to that of the two-dimensional lepidocrocite structure.

## ASSOCIATED CONTENT

Supporting information is available free of charge.

Electron energy loss spectra for one- and two-dimensional material with quantification results (inset). Simulated plan-view images of a single unit cell thick lepidocrocite $TiO_2$ for a range of applied tilt in x & y directions (deg.). Comparison of the edge energies for $TiO_2$ foils of different widths, with the edges oriented normal to the y- or x-directions. For edges normal to the y-axis, an edge configuration involving a structural reconstruction (filled circles) was found to be more stable than the unreconstructed configuration (empty circles). These reconstructed y-direction edges were considered when evaluating the effect of carbon on the edge energies in Fig. 5 of the main manuscript. Coordination number (CN), bond distance (R), the Debye-Waller factor ($\Delta\sigma^2$) information from the fitting results of the absolute Fourier transforms of extended fine structure region (k space) for two-dimensional $TiO_2$

## AUTHOR INFORMATION

**Corresponding author**

Per O.Å. Persson

Department of Physics, Chemistry and Biology (IFM), Linköping University, 58183 Linköping, Sweden

Wallenberg Initiative Materials Science for Sustainability (WISE), Linköping University, Department of Physics, Chemistry and Biology (IFM), 58183 Linköping, Sweden

per.persson@liu.se.

**AUTHOR CONTRIBUTION**

Eric Nestor Tseng -Investigation, Writing original draft

Jonas Björk-Formal analysis, Writing original draft

Risha Achaiah Iythichanda-Validation, Writing -review and editing

Wei Zheng-Investigation, Writing -review and editing

Jie Zhou-Investigation, Writing -review and editing

Hatim Alnoor-Investigation, Writing -review and editing

Wei Hsiang Huang-Investigation, Writing -review and editing

Ming-Hsien Lin-Investigation, Supervision, Writing -review and editing

Johanna Rosen-Supervision, Writing -review and editing

Per O.Å. Persson-Conceptualization, Supervision, Writing -review and editing

**FUNDING SOURCES**

The Swedish Research Council (2021-04499)

The Swedish Energy Agency for project grants (P2020-90149)

The Knut and Alice Wallenberg Foundation

The Swedish Foundation for Strategic Research (SSF)

ARTEMI

The Swedish National Infrastructure in Advanced Electron Microscopy (2021-00171 and RIF21-0026)

The Swedish Government Strategic Research Area (Faculty Grant SFO-Mat-LiU 2009-00971)

The Swedish Research Council (2022-06725).

**NOTES**

The authors declare no competing financial interest

**ACKNOWLEDGMENT**

The authors thank the Swedish Research Council (VR) and The Swedish Energy Agency for project grants (2021-04499, and P2020-90149), respectively. The Knut and Alice Wallenberg Foundation is acknowledged for support of the Linköping Electron Microscopy Laboratory. VR and the Swedish Foundation for Strategic Research (SSF) are further acknowledged for access to ARTEMI, the Swedish National Infrastructure in Advanced Electron Microscopy (2021-00171 and RIF21-0026). The authors also acknowledge the Swedish Government Strategic Research Area in Materials Science on Advanced Functional Materials at Linköping University (Faculty Grant SFO-Mat-LiU No. 2009-00971). The computational analyses were enabled by resources provided by the National Academic Infrastructure for Supercomputing in Sweden (NAISS), partially funded by the Swedish Research Council through grant agreement no. 2022-06725.

**REFERENCES**

1. Novoselov, K. S.; Geim, A. K.; Morozov, S. V.; Jiang, D.; Zhang, Y.; Dubonos, S. V.; Grigorieva, I. V.; Firsov, A. A. Electric Field Effect in Atomically Thin Carbon Films. *Science* **2004**, *306*, 666–669. https://doi.org/10.1126/science.1102896

2. Lin, Y.; Williams, T. V.; Connell, J. W. Soluble, Exfoliated Hexagonal Boron Nitride Nanosheets. *J. Phys. Chem. Lett.* **2010**, *1*, 277–283. https://doi.org/10.1021/jz9002108

3. Joensen, P.; Frindt, R. F.; Morrison, S. R. Single Layer $MoS_2$. Mater. Res. Bull. 1986, 21, 457–461. https://doi.org/10.1016/0025-5408(86)90011-5

4. Naguib, M.; Kurtoglu, M.; Presser, V.; Lu, J.; Niu, J.; Heon, M.; Hultman, L.; Gogotsi, Y.; Barsoum, M. W. Two-Dimensional Nanocrystals Produced by Exfoliation of $Ti_3AlC_2$. Adv. Mater. 2011, 23, 4248–4253. https://doi.org/10.1002/adma.201102306

5. Kumar, S. G.; Gomathi Devi, L. Review on Modified $TiO_2$ Photocatalysis under UV/Visible Light: Selected Results and Related Mechanisms on Interfacial Charge Carrier Transfer Dynamics. *J. Phys. Chem. A* **2011**, *115*, 13211–13241. https://doi.org/10.1021/jp204364a

6. Schaub, R.; Thostrup, P.; Lopez, N.; Lægsgaard, E.; Stensgaard, I.; Nørskov, J. K.; Besenbacher, F. Oxygen Vacancies as Active Sites for Water Dissociation on Rutile $TiO_2$ (110). *Phys. Rev.Lett.* **2001**, *87*, 266104. https://doi.org/10.1103/PhysRevLett.87.266104

7. Wang, A. S. D.; Sasaki, T. Titanium Oxide Nanosheets: Graphene Analogues with Versatile Functionalities. *Chem. Rev.* **2014**, *114*, 9455–9486. 10.1021/cr400627u

8. Bourikas, K.; Kordulis, C.; Lycourghiotis, A. Titanium Dioxide (Anatase and Rutile): Surface Chemistry, Liquid–Solid Interface Chemistry, and Scientific Synthesis of Supported Catalysts. *Chem. Rev.* **2014**, *114*, 9754–9823. https://doi.org/10.1021/cr300230q

9. Wang, X.; Li, Z.; Shi, J.; Yu, Y. One-Dimensional Titanium Dioxide Nanomaterials: Nanowires, Nanorods, and Nanobelts. *Chem. Rev.* **2014**, *114*, 9346–9384. https://doi.org/10.1021/cr400633s

10. Ramos-Delgado, N. A.; Gracia-Pinilla, M. A.; Mangalaraja, R. V.; O'Shea, K.; Dionysiou, D. D. Industrial Synthesis and Characterization of Nanophotocatalysts Materials: Titania. *Nanotechnol. Rev.* **2016**, *5*, 467–

479. https://doi.org/10.1515/ntrev-2016-0007

11. Fang, W.; Xing, M.; Zhang, J. Modifications on Reduced Titanium Dioxide Photocatalysts: A Review. *J. Photochem. Photobiol., C: Photochem. Rev.* **2017**, *32*, 21–39. https://doi.org/10.1016/j.jphotochemrev.2017.05.003

12. Saeed, M.; Muneer, M.; Akram, N. Photocatalysis: An Effective Tool for Photodegradation of Dyes A Review. *Environ. Sci. Pollut. Res.* **2021**, *28*, 1–19. https://doi.org/10.1007/s11356-021-16389-7

13. Sasaki, T.; Watanabe, M.; Michiue, Y.; Komatsu, Y.; Izumi, F.; Takenouchi, S. Preparation and Acid–Base Properties of a Protonated Titanate with the Lepidocrocite-like Layer. *Chem. Mater.* **1995**, *7*, 1001–1006. https://doi.org/10.1021/cm00053a029

14. Sasaki, T.; Watanabe, M. Semiconductor Nanosheet Crystallites of Quasi-$TiO_2$ and Their Optical Properties. *J. Phys. Chem. B* **1997**, *101*, 10159–10161. https://doi.org/10.1021/jp9727658

15. Zhou, W.; Umezawa, N.; Ma, R.; Sakai, N.; Ebina, Y.; Sano, K.; Liu, M.; Ishida, Y.; Aida, T.; Sasaki, T. Spontaneous Direct Band Gap, High Hole Mobility, and Huge Exciton Energy in Atomic-Thin $TiO_2$ Nanosheet. *Chem. Mater.* **2018**, *30*, 6449–6457. https://doi.org/10.1021/ACS.CHEMMATER.8B02792

16. Dahlqvist, M.; Zhou, J.; Persson, I.; Ahmed, B.; Lu, J.; Halim, J.; Tao, Q.; Palisaitis, J.; Thörnberg, J.; Helmer, P.; Hultman, L.; Persson, P. O. Å.; Rosen, J. Out-of-Plane Ordered Laminate Borides and Their 2D Ti-Based Derivative from Chemical Exfoliation. *Adv. Mater.* **2021**, *33*, 2008361. https://doi.org/10.1002/adma.202008361

17. Badr, H. O.; ElMelegy, T.; Carey, M.; Natu, V.; Hassig, M. Q.; Johnson, C.; Qian, Q.; Li, C. Y.; Kushnir, K.; ColinUlloa, E.; Titova, L. V.; Martin, J.; Grimm, R. L.; Pai, R.; Kalra, V.; Karmakar, A.; Liang, K.; Naguib, M.; Wilson, O.; Magenau, A. J. D.; Montazeri, K.; Zhu, Y.; Cheng, H.; Torita, T.; Koyanagi, M.; Yanagimachi, A.; Ouisse, T.; Barbier, M.; Wilhelm, F.; Rogalev, A.; Persson, P.; Rosen, J.; Hu, Y.-J.; Barsoum, M. W. BottomUp, Scalable Synthesis of Anatase NanofilamentBased TwoDimensional Titanium CarboOxide Flakes. *Mater. Today* **2022**, *54*, 8–17. https://doi.org/10.1016/j.mattod.2021.10.033

18. Kresse, G.; Furthmüller, J. Efficient Iterative Schemes for Ab Initio Total-Energy Calculations Using a Plane-Wave Basis Set. *Phys. Rev. B* **1996**, *54*, 11169–11186. https://doi.org/10.1103/PhysRevB.54.11169

19. Blöchl, P. E. Projector Augmented-Wave Method. *Phys. Rev. B* **1994**, *50*, 17953–17979. https://doi.org/10.1103/PhysRevB.50.17953

20. Perdew, J. P.; Burke, K.; Ernzerhof, M. Generalized Gradient Approximation Made Simple. *Phys. Rev. Lett.* **1996**, *77*, 3865–3868. https://doi.org/10.1103/PhysRevLett.77.3865

21. Paidi, V. K.; Lee, B. H.; Ahn, D.; Kim, K. J.; Kim, Y.; Hyeon, T.; et al. Oxygen-Vacancy-Driven Orbital Reconstruction at the Surface of $TiO_2$ Core–Shell Nanostructures. *Nano Lett.* **2021**, *21*, 7953–7959. https://doi.org/10.1021/acs.nanolett.1c01995

22. Stiller, M.; Ohldag, H.; Barzola-Quiquia, J.; Esquinazi, P. D.; Amelal, T.; Bundesmann, C.; et al. Titanium 3d Ferromagnetism with Perpendicular Anisotropy in Defective Anatase. *Phys. Rev. B* **2020**, *101*, 014412. https://doi.org/10.1103/PhysRevB.101.014412

23. Yoshiya, M.; Tanaka, I.; Kaneko, K.; Adachi, H. First-Principles Calculation of Chemical Shifts in ELNES/NEXAFS of Titanium Oxides. *J. Phys.: Condens. Matter* **1999**, *11*, 3217–3230. https://doi.org/10.1088/0953-8984/11/16/003

24. Rafferty, B.; Brown, L. Direct and Indirect Transitions in the Region of the Band Gap Using Electron-Energy-Loss Spectroscopy. *Phys. Rev. B* **1998**, *58*, 10326–10337. https://doi.org/10.1103/PhysRevB.58.10326

25. Wang, L.; Sasaki, T. Titanium Oxide Nanosheets: Graphene Analogues with Versatile Functionalities. *Chem. Rev.* **2014**, *114*, 9455–9486. https://doi.org/10.1021/cr400627u

26. Huang, W. H.; Su, W. N.; Chen, C. L.; Lin, C. J.; Haw, S. C.; Lee, J. F.; et al. Structural Evolution and Au Nanoparticles Enhanced Photocatalytic Activity of Sea-Urchin-Like $TiO_2$ Microspheres: An X-Ray Absorption Spectroscopy Study. *Appl. Surf. Sci.* **2021**, *562*, 150127. https://doi.org/10.1016/j.apsusc.2021.150127

27. Shin, S. I.; Go, A.; Kim, I. Y.; Lee, J. M.; Lee, Y.; Hwang, S. J. A Beneficial Role of Exfoliated Layered Metal Oxide Nanosheets in Optimizing the Electrocatalytic Activity and Pore Structure of Pt-Reduced Graphene Oxide Nanocomposites. *Energy Environ. Sci.* **2013**, *6*, 608–617. https://doi.org/10.1039/C2EE22739H